\documentclass[aps,pra,english,groupedaddress, 11pt]{revtex4-2}
\usepackage{graphicx}
\usepackage{subfigure}
\usepackage{amsmath}
\usepackage{amssymb}
\usepackage[font=small,labelfont=bf]{caption}
\usepackage{amsfonts}
\usepackage[margin=2cm]{geometry}
\usepackage{mathtools}
\usepackage{multirow}
\usepackage{array} % Include this in your preamble for better table styling
\usepackage{subfig}
\usepackage{subcaption}
\usepackage[labelfont=bf, labelsep=period, font=small]{caption}
\usepackage[utf8]{inputenc}
\usepackage{dcolumn}
\usepackage{seqsplit}
\usepackage{relsize}
\usepackage{mathrsfs}
\usepackage{calrsfs}
\date{\today}
\newcommand{\be}{\begin{equation}}
	\newcommand{\bea}{\begin{eqnarray}}
		\newcommand{\bc}{\begin{center}}            
			\newcommand{\ee}{\end{equation}}
		\newcommand{\eea}{\end{eqnarray}}
	\newcommand{\ec}{\end{center}}

\begin{document}
	\title{Optimal performance of thermoelectric devices
	with small external irreversibility}
	\author{Rajeshree Chakraborty}
	\email[e-mail: ]{mp18011@iisermohali.ac.in}
	\author{Ramandeep S. Johal} 
	\email[e-mail: ]{rsjohal@iisermohali.ac.in}
	\affiliation{Department of Physical Sciences,\\
		Indian Institute of Science Education and Research Mohali,\\
		Sector 81, S.A.S. Nagar,\\
		Manauli PO 140306, Punjab, India.}
	\begin{abstract}
	In the thermodynamic analysis of thermoelectric devices,
	typical irreversibilities are for the
	processes of finite-rate heat transfer, heat leak and Joule heating. Approximate analyses often focus on either internal or external irreversibility, obtaining well-known expressions for the efficiency at maximum power (EMP), such as the Curzon-Ahlborn value for endoreversible model and the Schmiedl-Seifert form for exoreversible model.
	Within the Constant Properties model,
	we  simultaneously incorporate internal  as well as external irreversibilities. We employ the approximation of a symmetric and  small external irreversibility (SEI), allowing a
	tractable expression for EMP that
	depends on three parameters i) the ratio of internal to external thermal conductance ii) the figure of merit of the thermoelectric material and
	iii) the ratio of hot and cold reservoir temperatures.
	We study limiting forms of this EMP and compare our
	framework with the exact model as well as with
	other irreversible models in finite-time thermodynamics,
	such as the minimally nonlinear model.
	In particular, we argue that the TEG in endoreversible approximation can be mapped to
	the mesoscopic model of Feynman's ratchet in the high temperatures regime,
	thus providing an alternative to the viewpoint in literature
	where the TEG in linear regime is mapped
	to an exoreversible case.
    Finally, extending our study to the thermoelectric refrigerator under similar assumptions as for the generator, we
	analyze the efficiency at the maximum cooling power.
\end{abstract}
\maketitle	
\section{Introduction}
A thermoelectric generator (TEG) serves as an paradigmatic model for a realistic heat engine incorporating both internal and external sources of irreversibility \cite{Ioffebook,Gordon1991,Apertet2012}.
The recent global emphasis on energy awareness has underscored the importance of unconventional methods for generating electrical energy. TEGs, which leverage the Seebeck effect,
are eco-friendly in this respect \cite{DiSalvo1999,Rowebook}.
They have garnered increasing attention due to such advantages as minimal noise, zero emissions and extended operational life \cite{Gomez2014}.
Low-temperature TEGs are particularly versatile, finding applications in diverse areas where waste heat is prevalent, such as industrial engineering \cite{Miao2023}, photovoltaic generation \cite{Lamba2016,Yusuf2022}, aviation \cite{Samson2010}, electronic devices \cite{Mohammed2021}, and internal combustion engines \cite{Fern2018}.  In recent years, considerable efforts have been made to identify thermoelectric materials with enhanced performance \cite{Poudel2008,Chung2000}. In addition to improving the figure of merit  \cite{Nemir2010,Littman1961,Majumdar2004,Shakouri2011,Muralidharan2012,Feng2018}, the system analysis and optimization of a TEG are crucial for developing high-performance thermoelectric systems
\cite{Snyder2003, Iyyappan2017,Riffat2003,Pei2011,Goupil2011,Goldsmid2010, Meng2014, Chen2015, LambaGA2018}. In this quest, a prevalent optimization technique adopted for TEG is the power output maximization,
with an associated interesting parameter known as the efficiency at maximum power (EMP) \cite{VandenBroeck2005,Esposito2009,Ouerdane2015,Moreau2012}.

In the context of finite-time thermodynamics, Curzon-Ahlborn efficiency given by \(\eta_{\text{CA}} = 1 - \sqrt{T_c/T_h}\) is significant, where \(T_{h}\) and \(T_{c}\) being the temperatures of the hot and cold reservoirs, respectively. This expression for EMP was derived for a specific class of heat engines known as endoreversible heat engines \cite{Novikov1958,Curzon1975,Agrawal1997} which assume that irreversibilities arise solely from finite heat transfer rates between the working substance and the heat reservoirs. A different expression for EMP was later derived for
the so-called exoreversible heat engines, which attribute irreversibilities to internal dissipations, such as friction and Joule heating, while assuming perfect thermal contacts. Schmiedl and Seifert \cite{SchmiedlSeifert2008} derived this efficiency as \(\eta_{\mathrm{SS}} = {\eta_{\mathrm{C}}}/({2 - \gamma \eta_{\mathrm{C}}})\), where \(\gamma\) ranges from 0 to 1 and takes the value \(1/2\) for homogeneous thermoelectric materials, indicating symmetric Joule heat dissipation in both hot and cold reservoirs \cite{Lu2014}. Kaur and Johal \cite{Jasleen2019} conducted a study on the optimal power of TEG with both internal and external irreversibilities characterized by a spatially dependent internal thermal conductivity which leads to asymmetric dumping of
Joule heat on the hot and cold sides. The inhomogeneity of internal thermal conductivity
was shown to enhance the performance of TEG-coupled hybrid systems such as
the molten carbonate fuel cell \cite{Chen2022}  and the evacuated U-tube
solar collector \cite{Yiwei2024}.
The effect of space-dependent electrical conductivity in a thermoelectric
cooler was also studied recently \cite{Hu2018, Ding2021}.

 The previous research on the optimization of TEGs indicates that a comprehensive and predictive  thermodynamic model incorporating both internal and external irreversibilities, is
 considered intractable i.e. the model cannot be solved analytically without introducing certain approximations.
 Indeed, a solvable model is crucial for gaining insight into operational regimes and a suitable device design.
The present work  analyzes the Constant Properties (CP) model
\cite{Ioffebook} by including internal irreversibilities and finite rates of heat transfer on both hot and cold
contacts---under the approximation of a small external irreversibility (SEI).
Optimizing the power output, we derive a compact general expression for EMP that depends on three quantities:
i) the ratio of cold to hot temperatures, ii) the figure of merit
of the thermoelectric material and iii) the ratio of internal to external thermal conductances. Different performance bounds are analyzed while recovering the known cases of endoreversible and exoreversible approximations.
We also study thermoelectric refrigerators (TER) under
the SEI approximation, where the focus is on the optimization of cooling power and to analyze the efficiency at maximum cooling power (EMCP). The bounds derived from this expression not only reveal the system's convergence to the well-established exoreversible model of TER \cite{Apertet2013}, but also extend the previous analysis.

Our paper is structured as follows. Section II details the model of a thermoelectric generator where in Section II.A we present the case of SEI and optimize the power output.
In Section II.B the properties of EMP are analyzed in some detail.
In Section II.C, we make comparison of our model with
the exact model using the experimental data on thermoelectric
materials. In Section III, we map
the endoreversible model of TEG with SEI to the Feynman's ratchet
in the high temperatures regime.
In Section IV, we present the model for a thermoelectric refrigerator and address the optimization of cooling power. Section V concludes with a summary and final remarks.
\section{Thermoelectric generator model}
\begin{figure}
     \centering
     \includegraphics[width=7cm]{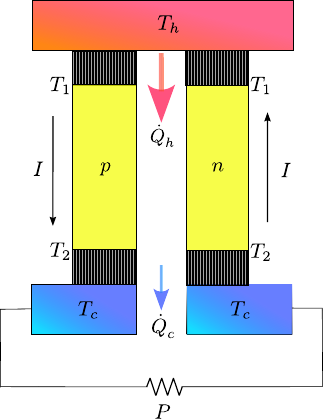}
     \caption{Schematic representation of a two-leg thermoelectric generator (TEG), illustrating the  heat source and the sink, the flow of heat and
     the extraction of electrical power via an external load.
     $p$- and $n$-type each represents a TEM, while
     the hatched segments
     are heat exchangers with symmetric thermal conductance $K$
     on hot and cold sides.}
     \label{model1}
 \end{figure}

 We consider the specific design of a two-leg configuration of TEG, as shown in
 Fig. \ref{model1}. The thermoelectric material (TEM) in n-type and p-type legs each is of
 length $L$ having areas of cross section $A_n$ and $A_p$ respectively.
 Within the CP model, the intrisic properties of the TEM
 are considered independent of the temperature or to be fairly fixed
 over the given range of temperatures.
 We denote electrical
 resistivities as $\rho_{n}$ and $\rho_{p}$, thermal conductivities
 as \(k_n\) and
 \(k_p\), and Seebeck coefficients  as \(\alpha_n\) and \(\alpha_p\), respectively. Then, the global intrinsic properties of
 the TEG are given as the  electrical resistance
 $R = (\rho_n /A_n + \rho_p /A_p)L$, thermal conductance
 $K_0 = (k_n A_n + k_p A_p)/L$ and Seebeck coefficient, $\alpha = \alpha_p-\alpha_n $ \cite{Pedersen2007}.
 Let $K_h$ and $K_c$ represent the thermal
 conductances of the heat exchangers that
 connect the TEM to the heat source  and the heat sink, respectively.
For analytic simplicity, we consider the symmetric case, $K_h=K_c\equiv K$,  
and small external
irreversibility (SEI) i.e. $K$ is large, but finite.
%It will be seen below, that operation of TEG requires that $K_0$ be small compared to $K$.
The case of symmetric $K$ may be understood in terms of a
simple design
constraint as follows. For the given TEM geometry, let $K_h = U_h A_h$
and $K_c = U_c A_c$, be defined as the product of the heat transfer 
coefficient $U_i$ and the area of the heat exchanger $A_i$, on both
hot and cold sides.
Under the Finite Physical Dimensions Thermodynamics 
\cite{Bejan1995, Feidt2012,JasleenPRE},
the physical dimensions of the heat exchangers are also
recognised as optimizable variables. A specific choice 
may be driven by the material costs or design constraints.
In the present case, a simple design may be where
the area of the heat exchanger
is equal to the area available on the themoelectric module i.e. 
$A_p + A_n = A$, on both hot and cold sides. Thus, we may set
$A_h = A_c = A$. Further, if we restrict to a single-material heat exchanger,
then $U_h = U_c = U$, 
and we obtain similar conductance of the heat
exchanger on either side: $K_h = K_c = K = U A$.

Now, let $I$ denote the constant value of the electric current flowing through the TEG. On the basis of the Onsager formalism and Domenicali's heat equation \cite{Pedersen2007,Domenicali1954}, the thermal currents at the hot and cold contacts
	of the TEM can be respectively written as
	\bea
	\dot{Q}_h^{} &=& \alpha T_{1} I +K_0 (T_{1}-T_{2})-\frac{1}{2}RI^{2}, 
	\label{hflux0}\\
	\dot{Q}_c^{} &=& \alpha T_{2} I +K_0(T_{1}-T_{2}) + \frac{1}{2} RI^{2} \label{cflux0},
	\eea
where  $T_1, T_2$ are the local temperatures at
the junction of TEM and heat exchangers, such that
$T_{1}<T_h$ $(T_{2}> T_c)$ (see Fig. 1).
The first term in the above equations corresponds to convective heat flow. The second term denotes the heat leakage across the TEM, while the last term accounts for the Joule heating effect as derived within the CP model.
We consider the heat transfer law
through the heat exchangers to be Newtonian so that the thermal currents
at the hot and cold junctions can be written as
 \begin{align}
\dot{Q}_{h} &= K(T_{h}^{}-T_{1}^{}),
\label{qhnew}\\
\dot{Q}_{c} & = K(T_{2}^{}-T_{c}^{}),
\label{qcnew}
\end{align}
 respectively.
In the limit when the thermal contacts between the reservoirs and TEM are ideal
($K\to \infty$), we have $T_1 \to T_h$ and $T_2 \to T_c$, and
the flux equations are simplified to:
	\bea
	\dot{Q}_h^{} &=& \alpha T_{h} I +K_0 (T_{h}-T_{c})-\frac{1}{2}RI^{2},
	\label{hflux}\\
	\dot{Q}_c^{} &=& \alpha T_{c} I +K_0(T_{h}-T_{c}) + \frac{1}{2} RI^{2} \label{cflux}.
	\eea
For the non-ideal thermal contacts,
the flux-matching conditions equate Eqs. (\ref{hflux0})
and (\ref{qhnew}) (similarly Eq. (\ref{cflux0}) equals (\ref{qcnew})), so that we
can explicitly solve for
$T_1$ and $T_2$:
\begin{equation}  \label{T1}
 T_{1}=\frac{-2K \alpha  T_h I  + (K + {2 K_0}) R I^2
- \alpha R I^{3} + 2K (K T_h + K_0(T_c+T_h))}{2[K(K+2K_0)-\alpha^2 I^2]},
\end{equation}
 \begin{equation} \label{T2}
     T_{2}=\frac{2K \alpha  T_c I  + (K + {2 K_0}) R I^2
+ \alpha R I^{3} + 2K (K T_c + K_0(T_c+T_h))}{2[K(K+2K_0)-\alpha^2 I^2]}.
 \end{equation}
Substituting these expressions back into Eqs. (\ref{hflux0}) and (\ref{cflux0}),
we can obtain explicit expressions for the hot and the cold fluxes. Then, the power output ($P=\dot{Q}_{h}-\dot{Q}_{c}$)
is obtained in the form:
\begin{equation}\label{Pexact}
    {P} = \frac{\alpha \left( T_h - T_c \right) K^2 I
    -K \left( K + 2 K_0 \right) R I^2 - \alpha^2 ( T_c + T_h) K I^2}{K \left( K + 2 K_0 \right)-\alpha^2 I^2 }.
\end{equation}
However, the solution for the optimal power  from
the above expression  hardly provides an insight.
Various approximations have been considered in literature, such as endoreversible approximation
which treats $R$ and $K_0$ to be negligible, or the exoreversible approximation
which implies $K\to \infty$ and $K_0\to 0$ limits. These studies aim to exclude either
internal or the external irreversibility. The so-called strong-coupling
approximation which neglects $K_0$ is also artificial since in dealing
with real materials, it is difficult to reduce $K_0$ without
increasing $R$, a stumbling block towards
achieving a high figure of merit [Eq. (\ref{figm})].
In this paper, we formulate our model with  both the internal and
external irreversibilities, however, with the simplification that
the latter irreversibility is assumed to
be small (as qualified below) as well as
symmetric on the two sides.

% The assumption of Newton's law requires that if $K$ is large then
% the temperature difference $T_h - T_1$ and $T_2 -T_c$ must be small.
% This

\subsection{Regime of small external
irreversibility (SEI) }
We define the regime of SEI as the parameter regime
where $1/K$ is regarded as a small parameter and we retain
terms in thermodynamic expressions only up to linear order in $1/K$.
Then, the thermal flux equations (leading to
Eq. (\ref{Pexact})) can be written in an
approximate form as:
\begin{align}
\dot{Q}_{h} =&\left(1-\frac{2 K_0}{K}\right)
\left[\alpha T_h I + K_0(T_h-T_c)\right]
%\nonumber  \\ &
-\left(\frac{R}{2}+\frac{{\alpha}^{2} T_{h}}{K}\right){I}^{2}+ \frac{\alpha R}{2 K} I^{3},
\label{hi3} \\
\dot{Q}_{c} =&\left(1-\frac{2 K_0}{K}\right)
\left[\alpha T_c I + K_0(T_h-T_c)\right]
%\nonumber\\ &
+ \left(\frac{R}{2}+\frac{{\alpha}^{2} T_{c}}{K}\right){I}^{2}+ \frac{\alpha R}{2 K} I^{3}.
\label{ci3}
\end{align}
The expression of power output is given by
\begin{equation}
P= \left(1-\frac{2 K_0}{K}\right) \alpha\left(T_h-T_c\right)I-\left(R+\frac{{\alpha^{2}}}{K}\left(T_{h}+T_{c}\right)\right)I^{2},
\label{pwr}
\end{equation}
where the cubic terms in $I$ cancel out, retaining a quadratic
expression for the power output.

%one may scale the fluxes by the factor $(1-2k)$.

Since the thermal fluxes
still contain the cubic term, the present model
is beyond the linear-irreversible regime \cite{Apertet2012}.
The performance of a thermoelectric device is usually
characterized by the figure of merit of TEM, defined as
\be
z= \frac{\alpha^2T_h}{R K_0}.
\label{figm}
\ee
In the case of SEI, apart from the parameter $z$,
the performance also depends on the parameter
\be
k = \frac{K_0}{ K}.
\ee
For instance, the positivity of power in Eq. (\ref{pwr}) requires that $k < 1/2$.

Now, the power output vanishes
when $I=0$, as well as in the short-circuit (sc)
limit, given by
\be
I_{\rm sc}=
\frac{\alpha (T_h-T_c)}{R}
\frac{(1-2k)}{1 +k z \left(2-\eta_{\rm C}^{}\right)}.
\ee
For a given TEM with a finite $z$ value, as $k\to 0$ (implying $K \gg K_0$),
the maximum current reduces to:
$i_{\rm sc}^{} = \alpha (T_h-T_c)/R$ \cite{Gordon1991}.
Clearly, the regime of a non-zero power output shrinks (since $I_{\rm sc} < i_{\rm sc}^{}$)
in the presence of external irreversibility.
Now, optimizing $P$ with respect to $I$
in the interval $[0,I_{\rm sc}^{}]$,
the optimal current is  $I^* = I_{\rm sc}/2$.
The expression for optimal power is given by
\be\label{Pmax}
 P^* = \frac{\alpha^2(1-2 k)^2 T_{h}^{2} \eta_{\rm C}^{2}}
 {4 R \left[1 +k z \left(2-\eta_{\rm C}^{}\right)\right]}.
\ee
In $k \to 0$ limit, we obtain
 $P^* = {\alpha^2  T_{h}^{2} \eta_{\rm C}^{2}}/
 {4 R }$ \cite{Gordon1991}.
Further, the difference in the junction temperatures,
 at the optimal power, is given by:
\begin{equation}
    T_{1}^ {*}-T_{2}^ {*} = \left(T_h-T_c \right) \left( 1-
    k\left[ z +2 -\frac{z\eta_{\rm C}^{}}{2} \right] \right).
\end{equation}
As expected, for $k\to 0$, the temperature difference between
the ends of TEM approaches
the difference in reservoir temperatures.
\subsection{Efficiency at maximum power (EMP)}
EMP is an important quantifier to
study of the performance of a finite-time engine.
Defined as the ratio of the optimal power output ($P^*$) to the
corresponding hot flux ($\dot{Q}_h (I^*)$), the EMP
of our model can be compactly
written as
\begin{equation}
 \eta^* = \frac{A}{B},
 \label{etab}
\end{equation}
where we denote
\begin{align}
A=& 4 (1-2 k)  \left(2 k z + 1-k z \eta_{\rm C}^{}\right)^2, \nonumber\\
B=& \left[4 k (-4 k z \eta_{\rm C}^{}+18 k z + k + 3 z + 10)+3\right] k z \eta_{\rm C}^{}
- 2(2 k z+1) \left[2 (14 k+5) k z+22 k+1\right] \nonumber\\
 & + \frac{8}{z \eta_{\rm C}^{}} (2 k z+1)^2 (2 k z+z+2) \nonumber,
\end{align}
which is expressed only in terms of the parameters
 $z$, $\eta_{\rm C}^{}$ and $k$.
 The above closed form expression for EMP quantifies the influence
of both internal and external irreversibilities in a TEG.
The known cases of only internal (or external)
irreversibility may be obtained from the above expression.
For a finite $z$ value and $k\to 0$,
we obtain
\begin{equation}
\lim_{k \rightarrow 0} \eta^* = \frac{2 z \eta_{\rm C}^{}}{8+ z(4-\eta_{\rm C}^{})}.
\label{eta8z}
% \eta^* = \frac{2 \eta_C}{4-\eta_C}.
\end{equation}
This expression is bounded from above by
the large-$z$ limit which yields the exoreversible model
($K = \infty$, $K_0=0$).
\begin{figure}
    \centering
    \includegraphics[width=11cm]{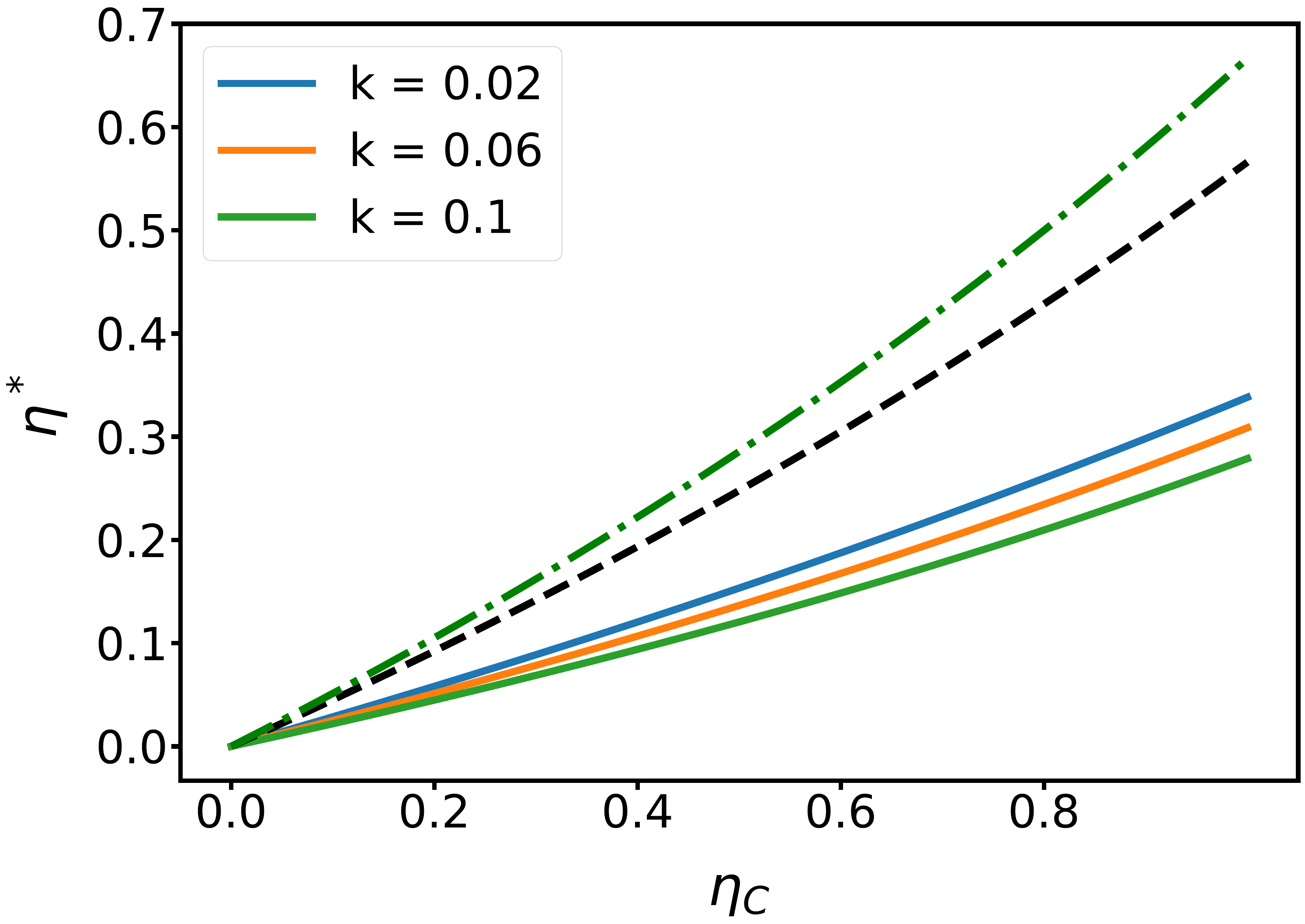}
    \caption{Efficiency at maximum power (EMP), Eq. (\ref{etab}), at a given $k<1/2$
    vs. Carnot efficiency ($\eta_{\rm C}^{}$) for $z=3$.
    With $k$ decreasing from
    bottom upwards, the EMP approaches the black dashed curve in the limit $k\to 0$,
    given by Eq. (\ref{eta8z}). The top dot-dashed curve, given by Eq. (\ref{effkz}), is approached in the ideal limit $z\to \infty$. }
    \label{eta_k}
\end{figure}
These two limits may be compactly depicted as:
  \begin{align}
  \lim_{z \rightarrow \infty}\lim_{k \rightarrow 0} \eta^*
    & =\frac{2 \eta_{\rm C}^{}}{4-\eta_{\rm C}^{}},
    \label{effkz}
    \end{align}
    which is $\eta_{\rm SS}$ for the symmetric case ($\gamma=1/2$)
    in the exoreversible approximation.
These limits are depicted in Fig. 2.

On the other hand,
first taking the $z\to \infty$ limit
leads to the endoreversible model. This can be
studied by setting $R=0$ and $K_0 =0$ in the flux
equations (\ref{hi3}) and (\ref{ci3}), which yields
\begin{align}
 \dot{Q}_{h} & =  \alpha T_h I- \frac{{\alpha^{2} T_h}}{K} I^{2},
 \label{qhK}\\
\dot{Q}_{c} & =  \alpha T_c I + \frac{{\alpha^{2}T_c}}{K} I^{2}.
\label{qcK}
\end{align}
Note that the above flux equations imply an endoreversible
model close to the reversible limit since we obtained
this limit under the approximation of SEI.
A more detailed discussion of the endoreversible
model is taken up in Ref. \cite{JasleenPRE}.
% which yield
% $P  =  \alpha\left(T_h-T_c\right)I- {{\alpha^{2}}}\left(T_{h}+T_{c}\right) I^{2}/K$.
The power output is now optimized at
$I = K (T_h -T_c)/[2\alpha(T_h+T_c)]$, yielding the EMP as
\begin{equation}
 \eta^* = \frac{2-\eta_{\rm C}^{}}{4-3 \eta_{\rm C}^{}}\eta_{\rm C}^{}.
\label{effzk}
 \end{equation}
 %Note that EMP is independent of the parameter $K$.
    The above form for EMP is obtained in coupled
    linear-irreversible engines \cite{RaiJohal2022}
    and, as we discuss in the next section, in
    Feynman's ratchet at high temperatures
    \cite{VSingh2018}.
    As a consistency check, we note that the above expression can also
be obtained from Eq. (\ref{etab}) by first taking
$z\to \infty$ followed by $k\to 0$.
Thus, we note that the limits of a vanishing-$k$ and a diverging-$z$
do not commute with each other. Actually, Eq. (\ref{effzk})
yields a higher value of EMP than CA value as well as Eq. (\ref{effkz}).
It may be noted that it is unrealistic to have
simultaneously zero values of electrical resistance as well
as thermal conductance for a thermoelectric.
Also, in view of the fact that the real thermoelectric materials
have a finite value of $z$, Eq. (\ref{effkz}) may be the more
relevant bound for TEGs.

The endoreversible limit [Eqs. (\ref{qhK}) and (\ref{qcK})]
of the above model
can be mapped to the minimally nonlinear irreversible model
\cite{Izumida2012, Izumida2015} in the strong coupling limit,
since the fluxes can be written in the form:
$ \dot{Q}_{h}  =  \alpha T_h I- \gamma_h I^{2}$ and
$\dot{Q}_{c}  =  \alpha T_c I + \gamma_c I^{2}$.
Here, the EMP is given by
\begin{equation}
 \eta_{\rm MN}^{} = \frac{\eta_{\rm C}^{}}{2}
 \left[1 - \frac{\eta_{\rm C}^{}}{2(1+{\gamma_c}/{\gamma_h}) } \right]^{-1}.
 \label{emn}
\end{equation}
where in the present case, ${\gamma_c}/{\gamma_h} = T_c/T_h$. Thereby,
Eq. (\ref{emn}) reduces to (\ref{effzk}).
Thus, our model with SEI generalizes the minimal 
model in the context of thermoelectric engines 
by including the cubic terms in the electric current.
% Note that whereas in the minimally nonlinear model, the themal
% fluxes are written up to quadratic term in the current,
% our model with SEI goes beyond it by considering the
% fluxes up to the cubic term.

To gain further insight into the general expression
for EMP, we consider
small temperature difference or  $\eta_{\rm C}^{} \ll 1$,
whereby EMP can be expressed
in series form as:
\be
 \eta^{*}=\frac{ (1-2 k )z}{2(2 k z+ z+ 2)} \eta_{\rm C}^{}
 +\frac{(4 k (1-3 k)+1) z^2 }{8 (2 k z+z+2)^2}
 \eta_{\rm C}^2 +{\cal O}\left(\eta_{\rm C}^3\right).
 \label{empexpandzk1}
\ee
For finite $z$ and $k\to 0$ [Eq. (\ref{eta8z})], the series is
simplified as follows.
\be
 \eta^{*}=\frac{z}{2( z+ 2)} {\eta_{\rm C}^{}}
 +\frac{z^2 }{ 8(z+2)^2}
 {\eta_{\rm C}^2} +{\cal O}\left(\eta_{\rm C}^3\right). \label{empexpandzk0}
\ee
Further,
in the limit of large $z$,
we obtain universal
terms in the series for EMP:
\be
\eta^*=\frac{\eta_{\rm C}^{}}{2}+\frac{\eta_{\rm C}^2}{8}+{\cal O}
\left(\eta_{\rm C}^{3}\right).
\label{unieff}
\ee
Interestingly, the same expansion is obtained up to the quadratic
term if we reverse the order of the two limits.
This shows an underlying aspect of the universality
of EMP beyond the linear response regime, which is independent
of the way the limit is approached. In general, the factor 1/2 in
the linear term [Eq. (\ref{unieff})]
represents the upper bound of
EMP in the linear response regime \cite{VandenBroeck2005}
while the 1/8 factor in the quadratic term
may be related to a certain left-right symmetry in
the system \cite{Esposito2009}.
Finally, it is interesting to compare the third order
term in the series expansion of Eqs. (\ref{effkz})
and (\ref{effzk}) which are respectively given as:
\be
\eta^*=\frac{\eta_{\rm C}^{}}{2}+\frac{\eta_{\rm C}^2}{8}+
\frac{1}{32}\eta_{\rm C}^{3} +
{\cal O}
\left(\eta_{\rm C}^{4}\right).
\label{SS3}
\ee
\be
\eta^*=\frac{\eta_{\rm C}^{}}{2}+\frac{\eta_{\rm C}^2}{8}+
\frac{3}{32}\eta_{\rm C}^{3} +
{\cal O}
\left(\eta_{\rm C}^{4}\right).
\label{FR3}
\ee
We note that up to third order in $\eta_{\rm C}^{}$, these two efficiencies lie symmetrically
above and below the CA value, which is given as:
\be
\eta_{\rm CA}^{} =\frac{\eta_{\rm C}^{}}{2}+\frac{\eta_{\rm C}^2}{8}+
\frac{2}{32}\eta_{\rm C}^{3} +
{\cal O}
\left(\eta_{\rm C}^{4}\right).
\label{CA3}
\ee

\subsection{Comparison with the exact CP model}
We have compared the SEI model with the exact model using
 experimental values for the various parameters,
 as extracted from Ref. \cite{Hogblom2014} and
given in Table \ref{tab:1}.
In Fig. 3, parametric loop curves are drawn between
the power output and the corresponding efficiency.
The SEI model shows similar trend as far as the general
shape of the curves is concerned. The area enclosed
by the loops increases as the temperature gradient increses,
thus increasing both the maximum power and the maximum efficiency.
We expect a better agreement
of SEI model with the exact model for smaller $k$ values which is apparent in Fig. 3.
\begin{table}[htbp]
    \centering
    \renewcommand{\arraystretch}{1.} % Adjust row height for better readability
    \setlength{\tabcolsep}{8pt} % Adjust column spacing

    \begin{tabular}{|>{\centering\arraybackslash}p{0.8cm}||c|c|c|c|c|c|c|}
        \hline
        \textbf{Case} & \boldmath$T_h$ (K) & \boldmath$T_c$ (K) & 
        \boldmath$\eta_{\rm C}=1-\frac{T_c}{T_h}$ & 
        \boldmath$K_0$ (W K\textsuperscript{-1}) & \boldmath$\alpha$ (V K\textsuperscript{-1}) & \boldmath$R$ (\boldmath$\Omega$) &
        \boldmath$z = \frac{\alpha^2 T_h}{RK_0}$ \\
        \hline\hline
        1 & 384 & 324 & 0.16 & $8.6 \times 10^{-3}$ & $5.0 \times 10^{-4}$ & $7.7 \times 10^{-3}$ & 1.45 \\
        2 & 413 & 333 & 0.19 & $8.8 \times 10^{-3}$ & $5.0 \times 10^{-4}$ & $8.5 \times 10^{-3}$ & 1.38 \\
        3 & 444 & 342 & 0.23 & $9.0 \times 10^{-3}$ & $5.1 \times 10^{-4}$ & $9.2 \times 10^{-3}$ & 1.4 \\
        4 & 498 & 348 & 0.30 & $10.0 \times 10^{-3}$ & $5.0 \times 10^{-4}$ & $9.9 \times 10^{-3}$ & 1.26 \\
        \hline
    \end{tabular}

    \caption{Parameters for TEG used in Fig. \ref{plot4} and based on Ref. \cite{Hogblom2014}.}
    \label{tab:1}
\end{table}
\begin{figure}[htbp] % Floating environment
    \centering
    % First row of plots
    \begin{subfigure}
        \centering
        \includegraphics[width=8cm]{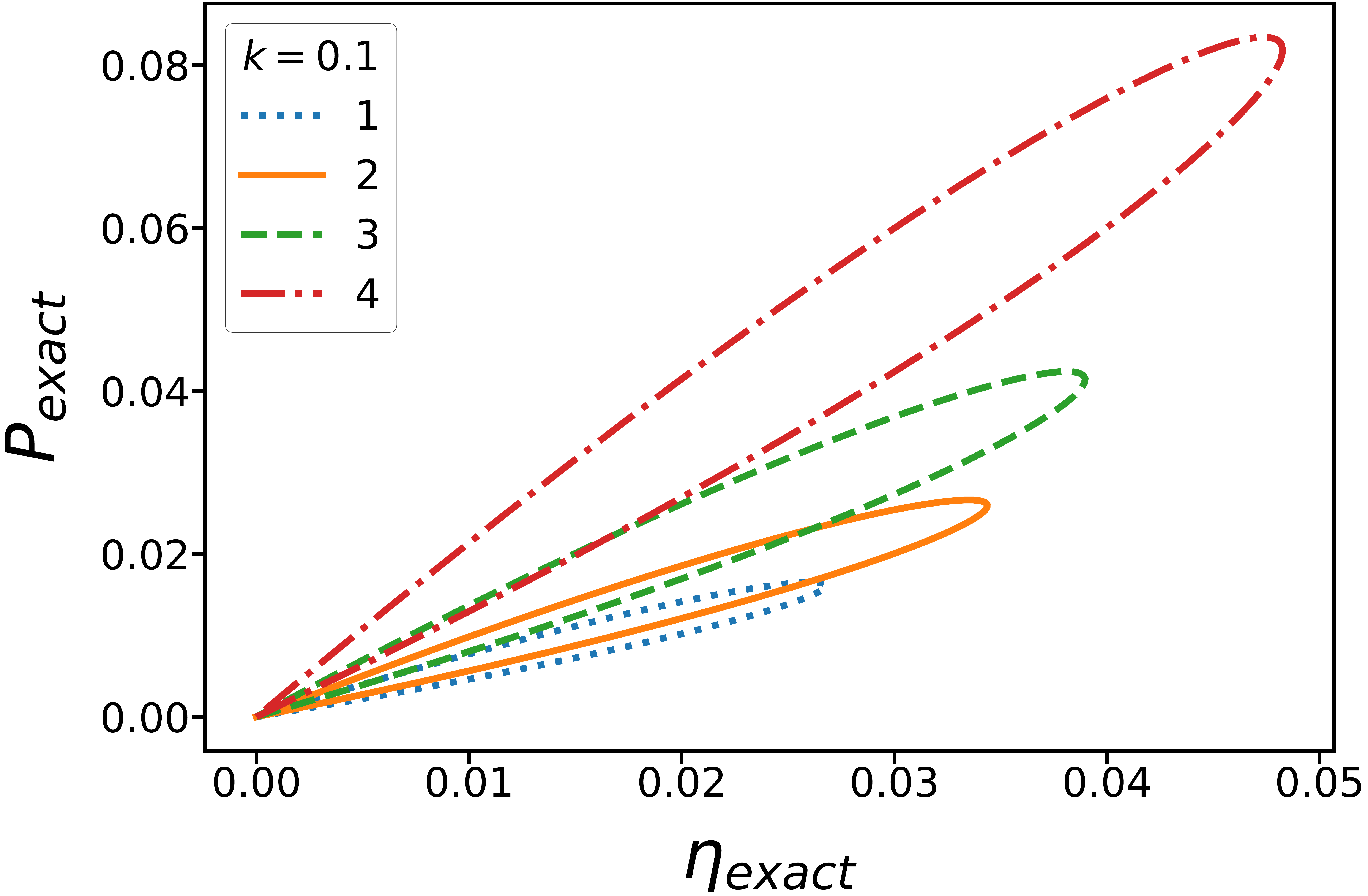}
        %\caption*{(a)}
    \end{subfigure}
    \hfill
    \begin{subfigure}
        \centering
        \includegraphics[width=8cm]{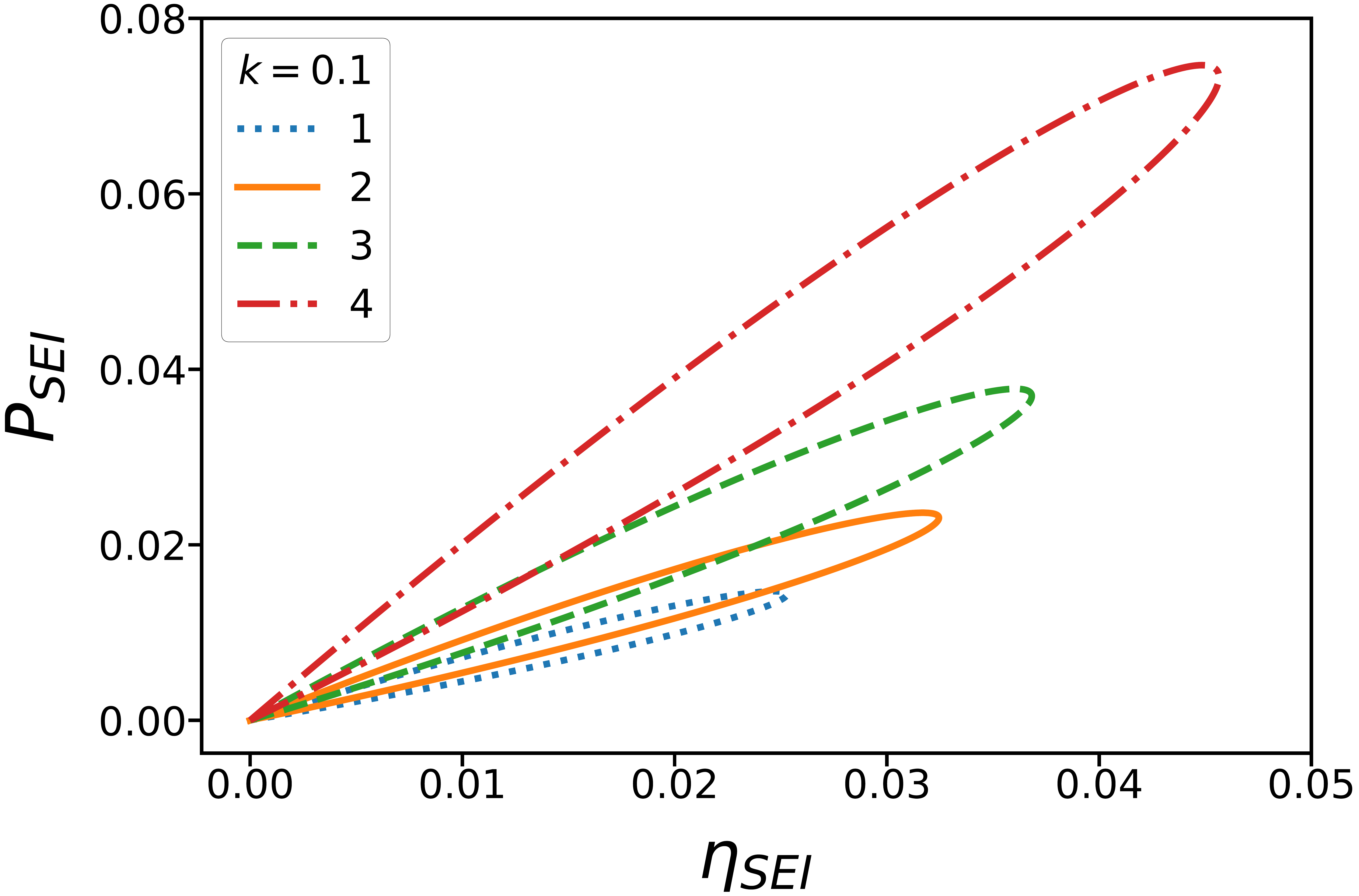}
        %\caption*{(b)}
    \end{subfigure}

    % Second row of plots
    \vspace{2em} % Adjust vertical spacing between rows if necessary
    \begin{subfigure}
        \centering
        \includegraphics[width=8cm]{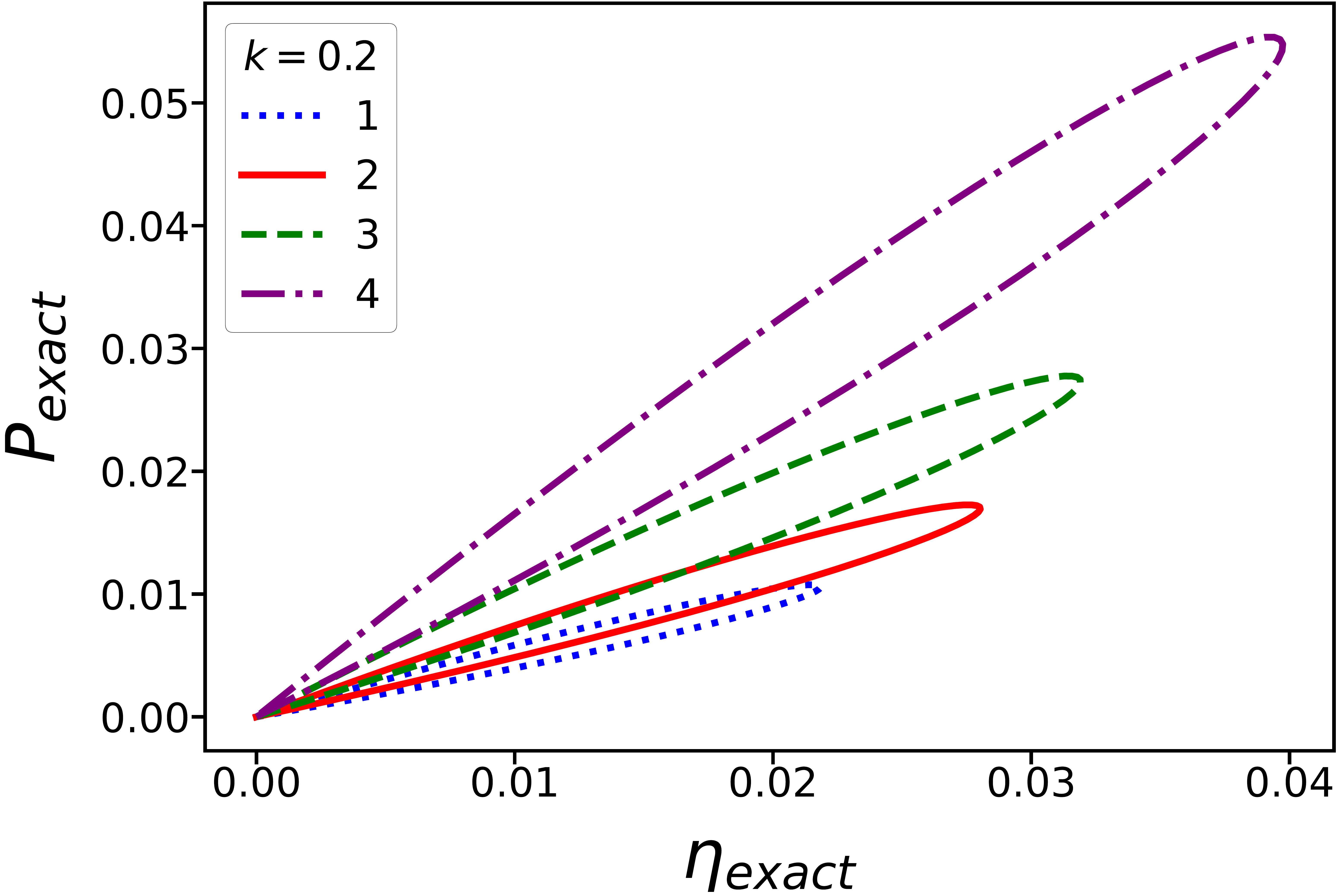}
        %\caption*{(c)}
    \end{subfigure}
    \hfill
    \begin{subfigure}
        \centering
        \includegraphics[width=8cm]{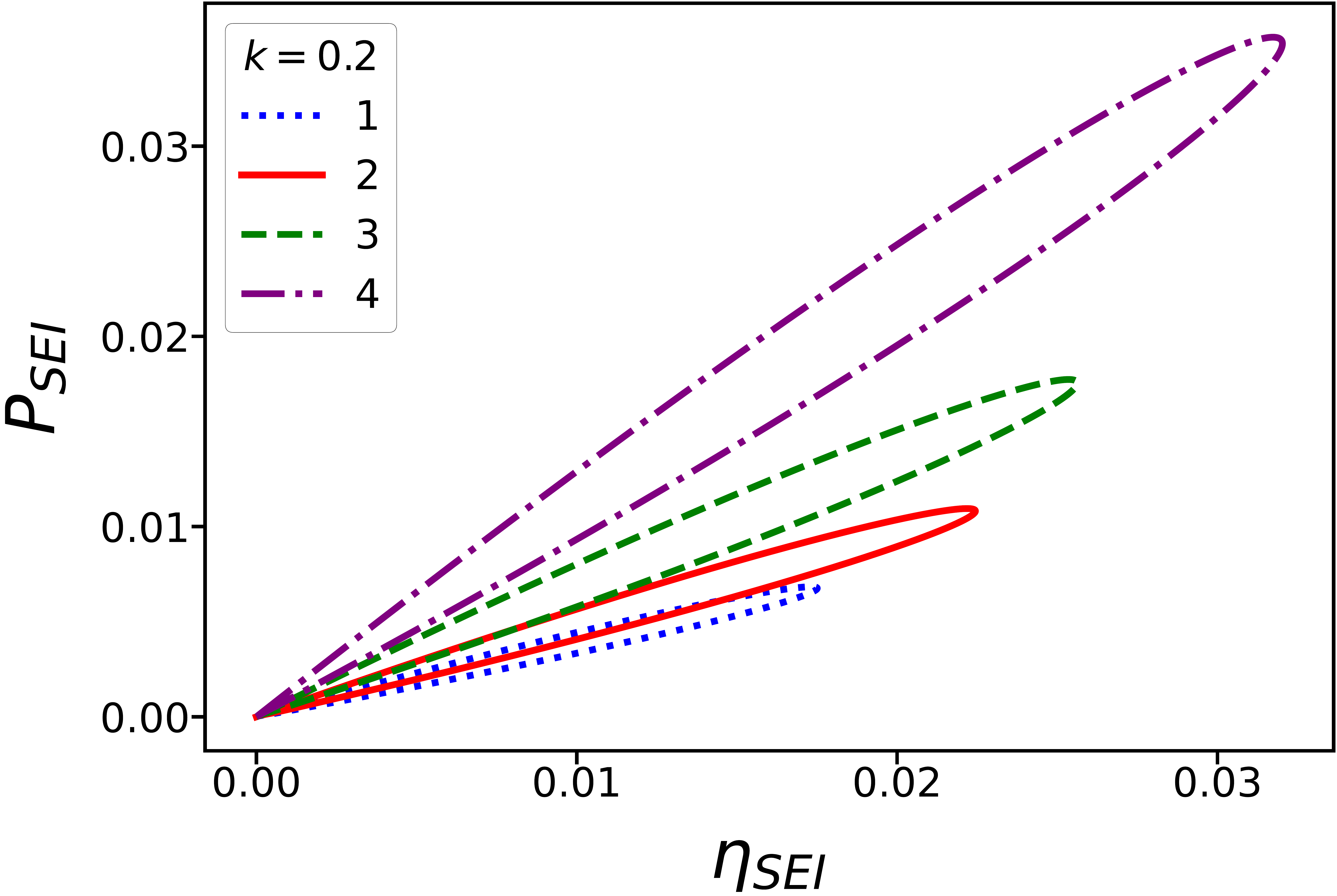}
        %\caption*{(d)}
    \end{subfigure}
    % Common caption for all plots
    \caption{Power output versus efficiency for the exact and the SEI models with \( k = 0.1 \) (top left and right) and \( k = 0.2 \) (bottom left and right). Legends 1, 2, 3 and 4 correspond to the data for the specific curve as listed in Table \ref{tab:1}. }
    \label{plot4}
\end{figure}

\section{Comparison with Feynman's ratchet model}
Feynman's ratchet and pawl \cite{Feynman} is a paradigmatic model for
the study of rectification of thermal fluctuations
in the mesoscopic regime and so serves as reference
model for the study of brownian or molecular
motors operating in living organisms.
The  model  consists
of a vane, immersed in a hot reservoir at temperature $T_h$,
and connected through an axle with a ratchet in contact with a
cold reservoir at $T_c$. Half-way along the axle, there is a wheel
from which a load is suspended. Because of the collisions of gas molecules,
the vane is subjected to thermal fluctuations, but
the ratchet is restricted to rotate in only one direction (forward)
due to the pawl which in turn is connected to a spring.
Let ${\epsilon_{2}^{}}>0$ be the amount of energy required
to overcome the elastic energy of the spring. Suppose that in each
step, the wheel rotates an angle $\phi$ and the torque
induced by the load is $L$. Then the system requires a
minimum of ${\epsilon_{1}^{}}={\epsilon_{2}^{}}+L\phi >0$ energy to lift the load that hangs from the axle. One part of the energy ${\epsilon_{1}^{}}$ is converted
into work $L\phi$ and the other is transferred as heat ${\epsilon_{2}^{}}$
to the cold reservoir.
Hence the rate of forward jumps of the ratchet is given by
$R_{\rm for}^{}=r_0 e^{-{\epsilon_{1}^{}}/T_h}$,
where $r_0$ is a rate constant assumed to be symmetric
with respect to the hot and cold ends. Also, we measure temperature
in the units of energy.
Similarly, the rate of the reverse jumps is $R_{\rm rev}^{}=r_0e^{-{\epsilon_{2}^{}}/T_c}$.
If $R_{\rm for}^{} > R_{\rm rev}^{}$, this system works as a steady-state
heat engine. The rate at which heat is exchanged with
the hot and the cold reservoir respectively is given by
\begin{equation}
\dot{Q}_h=r_0{\epsilon_{1}^{}} \left(e^{-{\epsilon_{1}^{}}/T_h}-
e^{-{\epsilon_{2}^{}}/T_c}\right) >0,
\label{q1dot}
\end{equation}
\begin{equation}
\dot{Q}_c=r_0{\epsilon_{2}^{}} \left(e^{-{\epsilon_{1}^{}}/T_h}-
e^{-{\epsilon_{2}^{}}/T_c}\right) >0.
\label{q2dot}
\end{equation}
The power output of the Feynman model is given by:
\begin{equation}
P_{\rm FR}^{} = \dot{Q}_h - \dot{Q}_c= r_0({\epsilon_{1}^{}} -  {\epsilon_{2}^{}})
\left(e^{-{\epsilon_{1}^{}}/T_h}- e^{-{\epsilon_{2}^{}}/T_c}\right).
\label{power}
\end{equation}
Since, we have  $\epsilon_{1}^{} > \epsilon_{2}^{}$ by construction,  the positivity of
the fluxes implies: ${{\epsilon_{2}^{}}}/{T_c} > {{\epsilon_{1}^{}}}/{T_h}$.
%
% \begin{equation*}
% \frac{{\epsilon_{2}^{}}}{T_2} > \frac{{\epsilon_{1}^{}}}{T_1}.
% \end{equation*}
The efficiency is given by
$\eta_{\rm FR}^{} =   P_{\rm FR}^{} /\dot{Q}_h =
1-\epsilon_{2}^{}/\epsilon_{1}^{}$.

The optimal
performance and EMP of this model have been
well studied \cite{Velasco2001, Tu2008,Apertet_FR2014, Thomas2015}.
Here, we work in the high temperatures regime where the energy scales
 are still small compared to the temperatures
of the reservoirs, i.e. $\epsilon_1 \ll T_h$ and $\epsilon_2 \ll T_c$. 
The linear regime was studied
in Ref. \cite{Apertet_FR2014} by expanding the exponentials
in the above expressions up to the first order and then
Feynman's model was mapped to the
exoreversible limit ($K \to \infty$ and $K_0 \to 0$)
of a thermoelectric generator. 
The analog of electric current $I$  in a thermoelectric
converter may be identified with the effective jump frequency
$\dot{N}_{\rm eff} =
R_{\rm for}^{} - R_{\rm rev}^{}$, under linear approximation.

We investigate the mapping between 
the Feynman ratchet and  TEG in the nonlinear regime by retaining
terms up to the second order in the exponentials \cite{VSingh2018}.
Interestingly, we find that Feynman model in
this limit can be mapped to a TEG within
endoreversible approximation.
Thus, we approximate the power output as
\be
P_{\rm FR}^{}(\epsilon_1,\epsilon_2) \approx r_0 
(\epsilon_1-\epsilon_2)\left(\frac{\epsilon_2}{T_c}
-\frac{\epsilon_1}{T_h}+\frac{\epsilon_1^2}{2 T_h^2}
-\frac{\epsilon_2^2}{2T_c^2}\right).
\label{PFRhot}
\ee 
The effective jump frequency in this regime is given by:
\be 
\dot{N}_{\rm eff} = r_0\left(\frac{\epsilon_2}{T_c}
-\frac{\epsilon_1}{T_h}+\frac{\epsilon_1^2}{2 T_h^2}
-\frac{\epsilon_2^2}{2T_c^2}\right).
\label{effjump}
\ee
Now, the above power output may be optimized
over both $\epsilon_1$ and $\epsilon_2$ \cite{VSingh2018}.
Interestingly, the EMP in this case is also 
given by Eq. (\ref{effzk}). So, can a similarity be 
established between the TEG and the ratchet
model in the high temperatures regime based on this observation?
However, note 
that the one-dimensional TEG model, under a given 
design and geometry, has only one optimization 
variable---the current $I$, whereas the Feynman's model involves
two variables, for instance $\epsilon_1$ and $\epsilon_2$.
In order to make a
reasonable comparison with the TEG model, 
we optimize the ratchet power  
over one of the variables, say $\epsilon_1$, 
and obtain the optimal power at a given efficiency as 
\be
P_{\rm FR}^{*}(\eta_{\rm FR}^{}) = \frac{16r_0 T_hT_c}{27}
\frac{(T_h(1-\eta_{\rm FR}^{})-T_c) \eta_{\rm FR}^{}}
{[(T_h(1- \eta_{\rm FR}^{})+T_c]^2},
\label{PFR_eta}
\ee
% Now, we intend to express the power output of the ratchet [Eq. (\ref{PFRhot})]
% in terms of its efficiency $\eta_{\rm FR}^{} = 1-\epsilon_2 /\epsilon_1$.
% For this purpose, we express the power
% as $P_{\rm FR}^{}(\eta_{\rm FR}^{}, \epsilon_1)$
% i.e. as a function of $\eta_{\rm FR}^{}$ and $\epsilon_1$.
% It is  remarkable that the optimum power, at
% a given efficiency $\eta_{\rm FR}^{}$, has a close semblence to
% the poer output of TEG model in endoreversible approximation.
% To observe this, we optimize the power
% with respect to $\epsilon_1$ by setting
% $(\partial /\partial \epsilon_1)P_{\rm FR}^{} = 0$,
the optimal value of $\epsilon_{1}^{}$ being given by 
\be
\epsilon_{1}^{*}= \frac{4T_h T_c}{3(T_h(1- \eta_{\rm FR}^{})
+ T_c)}.
\ee
Now, we would like to cast the TEG model, under endoreversible approximation, 
also in terms of its efficiency $\eta = 1-{\dot{Q}_{c}}/{\dot{Q}_{h}}$.
Using Eqs. (\ref{qhK}) and (\ref{qcK}),
the current $I$ can be written as a monotonic function 
%\cite{Jasleen2024}
of $\eta$ as 
\be
I= \frac{K}{\alpha}
\frac{{T_h}(1-\eta)-{T_c}}{{T_h}(1-\eta) + {T_c}},
\label{ieff}
\ee
making it possible to express 
the power output for a given efficiency
$P(\eta)= \eta \dot{Q}_{h}$,
in the endoreversible approximation, as:
\be
P(\eta)= 2 K T_h T_c \frac{({T_h}(1-\eta)-{T_c}) \eta }
{[T_h(1- \eta)+ T_c]^2}.
\label{Pteg_eta}
\ee
% In our endoreversible model of TEG with the Newton's law of heat transfer, the power output is given by $P=\dot{Q}_{h}^{}-\dot{Q}_{c}^{}$ where thermal currents are given by Eqs. (\ref{Qh}) and (\ref{Qc}). Using the relation, $\eta = 1- {\dot{Q}_{c}}/{\dot{Q}_{h}}$,
%
Note the formal similarity between the above expression
 and Eq. (\ref{PFR_eta}), thus connecting  two different models and
suggesting an analogous role for the parameters $r_0$ and $K$.
It is now clear that the EMP in either case is
given by the same expression, Eq. (\ref{effzk}).
Thus, the endoreversible
limit of our model can be looked upon either in terms of
the minimal nonlinear model (see Section II.B) or be
regarded analogous to Feynman's ratchet in the
high-temperatures, nonlinear regime.

% Eq. (\ref{PFR_eta}) can be expressed as
% $P_{\rm FR}^{} = \eta_{\rm FR}^{} \epsilon_{1}^{*} 
% \dot{N}_{\rm eff}^{*}$, where  

Further, we note that the effective jump 
frequency, at the optimal point $\epsilon_{1}^{*}$,
is given by 
\be
\dot{N}_{\rm eff}^{*} = \frac{4r_0}{9} 
\frac{{T_h}(1-\eta_{\rm FR}^{})-{T_c} }
{T_h(1- \eta_{\rm FR}^{})+ T_c}. 
\ee
Thus, we may infer that the effective jump frequency is
analogous 
to the quantity  $\alpha I$ [Eq. (\ref{ieff})] 
in the TEG model, with $r_0$ and $K$ playing similar
roles. A more detailed comparison between TEG and Feynman's
ratchet is beyond the scope of the present paper. 
%
% More importantly, we note from 
% Eq. (\ref{Neff}) that $\dot{N}_{\rm eff}^{}$
% vanishes only for Carnot efficiency
% ($\epsilon_2/\epsilon_1 = T_c/T_h$).  
%
\section{Thermoelectric refrigerator (TER)}
We now study the optimal performance of a TER
within the regime of SEI. Fig. 4 presents a schematic of the
TER showing the various fluxes and temperatures.
Heat flux from the cold reservoir is pushed against
the temperature gradient  using the external electric power,
due to thermoelectric effect. Note that for TER, $T_1 > T_h$
and $T_2 < T_c$.
Our target function here is the cold flux $\dot{Q}_c$, alos known as  the cooling power.
It has been noted \cite{Apertet2013} that an endoreversible model
of TER does not yield an optimum cooling power unlike the
exoreversible model.
 \begin{figure}
     \centering
     \includegraphics[width=7cm]{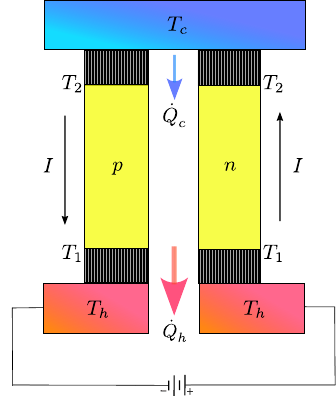}
     \caption{Schematic of a two-leg thermoelectric refrigerator, highlighting the key components such as the thermoelectric module, thermal contacts, heat fluxes and the power source.}
     \label{model}
 \end{figure}
Here, we investigate the effect of both external and internal
irreversibilities in the regime of SEI.
Within the CP model for a TER, the
thermal fluxes are given by
\bea
	\dot{Q}_h^{} &=& \alpha T_{1} I - K_0 (T_{1}-T_{2})+\frac{1}{2}RI^{2}, 
	\label{hfluxt}\\
	\dot{Q}_c^{} &=& \alpha T_{2} I - K_0(T_{1}-T_{2}) - \frac{1}{2} RI^{2}. \label{cfluxt}
	\eea
Assuming that the thermal flux through a heat exchanger
is Newtonian, the thermal currents
at the junctions are
\begin{align}
\dot{Q}_{h} &= K(T_{1}^{}-T_{h}^{}),
\label{qhnewt}\\
\dot{Q}_{c} & = - K(T_{2}^{}-T_{c}^{}).
\label{qcnewt}
\end{align}
From the above set of equations,
we can find here the explicit expresssions
for $T_1$ and $T_2$, similar to the case of TEG.
Upon substituting these into Eq. (\ref{hfluxt})
and (\ref{cfluxt}), we obtain the final heat
flux equations.

\subsection{Regime of  $\rm{SEI}$}
\begin{figure}
    \centering
    \includegraphics[width=11cm]{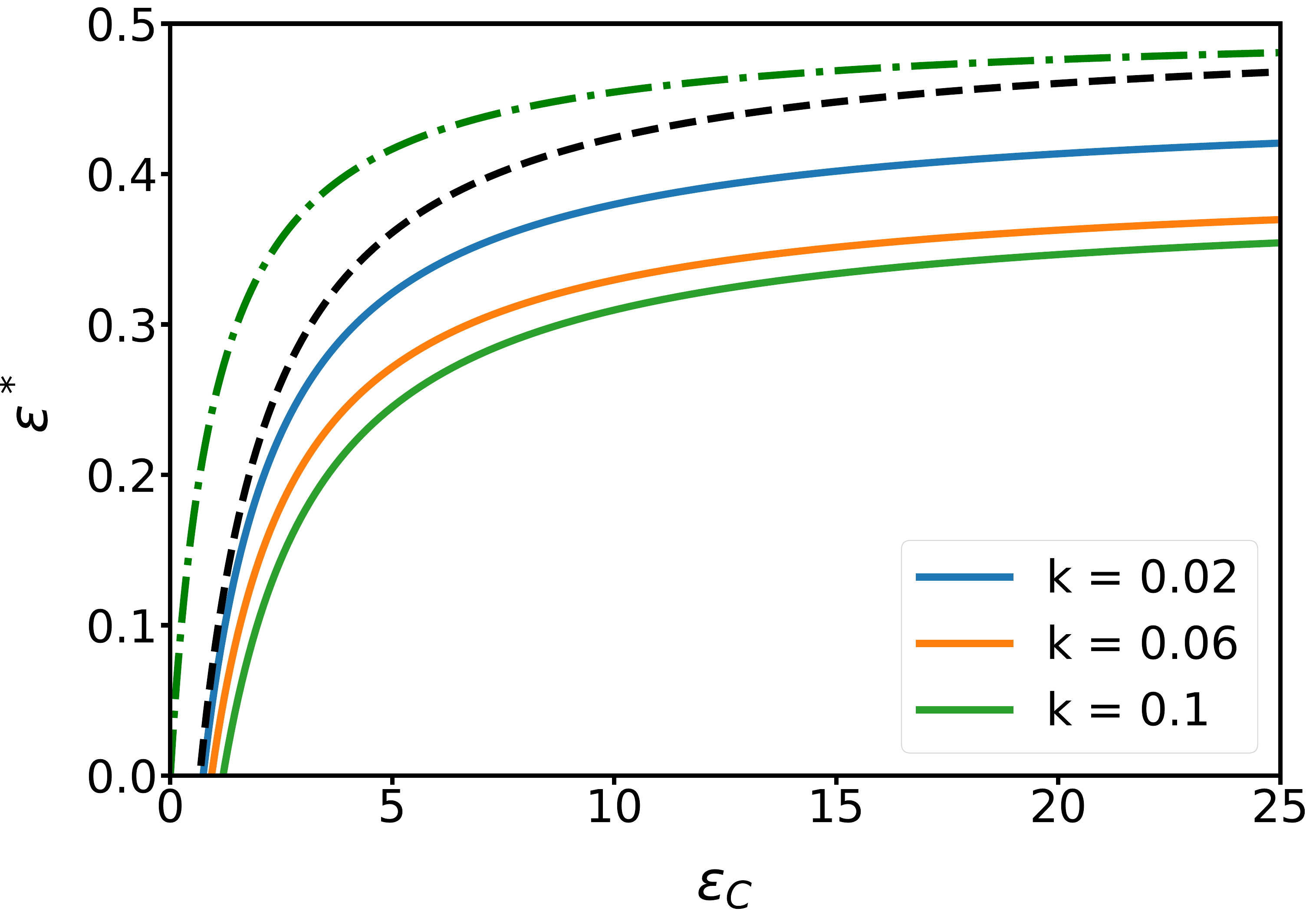}
    \caption{EMCP at  given $k$ values
    vs. $\epsilon_{\rm C}^{}$.
    As $k$ vanishes from
    bottom upwards, the EMCP approaches the dashed curve in the limit $k\to 0$,
    given by Eq. (\ref{epsk0}). $z=3$ is chosen in this figure.
    The top dot-dashed curve, given by $\epsilon_{\rm C}/2\left(1+\epsilon_{\rm C}\right)$, is reached for $z\to \infty$. }
    \label{eta_z}
\end{figure}

In the large-$K$
approximation, we truncate the final expressions for the heat fluxes keeping
terms in $1/K$, and obtain

\begin{align}
\dot{Q}_{h} =&\left(1-\frac{2 K_0}{K}\right)
\left[\alpha T_h I - K_0(T_h-T_c)\right]
%\nonumber  \\ &
+\left(\frac{R}{2}+\frac{{\alpha}^{2} T_{h}}{K}\right){I}^{2}+ \frac{\alpha R}{2 K} I^{3},
\label{hi4} \\
\dot{Q}_{c} =&\left(1-\frac{2 K_0}{K}\right)
\left[\alpha T_c I - K_0(T_h-T_c)\right]
%\nonumber\\ &
- \left(\frac{R}{2}+\frac{{\alpha}^{2} T_{c}}{K}\right){I}^{2}+ \frac{\alpha R}{2 K} I^{3}.
\label{ci4}
\end{align}
The power input ($P =\dot{Q}_{h}- \dot{Q}_{c}$) is given by
\begin{equation}
P= \left(1-\frac{2 K_0}{K}\right) \alpha\left(T_h-T_c\right)I+\left(R+\frac{{\alpha^{2}}}{K}\left(T_{h}+T_{c}\right)\right)I^{2}.
\label{Pwr}\\
\end{equation}
To optimize the cooling power [Eq. (\ref{ci4})], we set
$(\partial /\partial I)\dot{Q}_{c} = 0$.  The optimal
input current is
\be
I^*=\frac{K_0}{3\alpha k}(1+2 k \mathtt{z}-\sqrt{1-2 k  \mathtt{z}(1-2 k(3+ \mathtt{z}))},
\ee
where we define the figure
of merit in the refrigerator mode as
\begin{equation}
  \mathtt{z} = \frac{\alpha^2T_c}{R K_0}.
  \label{ztt}
\end{equation}
The condition $I^* >0$ requires that $k < 1/2$.
The condition of maximum, $(\partial /\partial I)^2\dot{Q}_{c}|_{I=I^*}  < 0$
is also verified.
For a given TEM, the limit $k = K_0/K \to 0$ yields
$I^* = \alpha T_c/R$. Likewise,
% with the strong coupling condition ($K_0 \rightarrow 0$) and
% reversible external contacts ($K \rightarrow \infty$), the
% optimal cooling power approaches
% $\dot{Q}_c^* \to {\alpha^2 T_c^2}/{2 R}$.
the optimal cooling power is given by:
\begin{equation}
 \lim_{k \rightarrow 0} \dot{Q}_c^* = K_0 T_c \left(\frac{\mathtt{z}}{2} -
 \frac{1}{\epsilon_{\rm C}} \right),
 \label{qc0}
\end{equation}
where $\epsilon_{\rm C}^{} = T_{c}/(T_{h} - T_{c})$
defines the  coefficient of performance
of a Carnot refrigerator.
Thus, the optimal cooling power condition requires that
$\mathtt{z} > 2/ \epsilon_{\rm C}$ if the external thermal contacts
are regarded as reversible. The corresponding
efficiency at maximum cooling power (EMCP),
defined as $\epsilon^* = {\dot{Q}_c^*}/{P(I^*)}$,
is given by:
\be
\lim_{k \rightarrow 0} \epsilon^* =\frac{\mathtt{z} \epsilon_{\rm C}-2}{2 \mathtt{z}\left(1+\epsilon_{\rm C}\right)}.
\label{epsk0}
\ee
For large-$\mathtt{z}$ values, it reduces monotonically  to the expression
$\epsilon_{\rm C}/2\left(1+\epsilon_{\rm C}\right)$,
 which is also the EMCP in  exoreversible limit \cite{Apertet2013}.
The general formula for EMCP in the present approximation ($0< k < 1/2$) depends in a complicated way on $k$, $\mathtt{z}$ and $\epsilon_{\rm C}$,
 which we do not present here for the sake of brevity and depict graphically in Fig. \ref{eta_z}.
\section{Conclusions}
In this work, we studied the 
performance of thermoelectric devices with 
an analytic model incorporating both internal and external irreversibilities.
While the real thermoelectric
materials have a finite figure of merit leading to
internal irreversibilities in the form of heat leak
and Joule heating, the external irreversibilities owing to
finite thermal conductance of the heat exchangers
also decrease the performance of the devices.
With the assumption of constant material properties,
thermoelectric models can be formulated in a simple, analytic 
form whereby the interplay of various irreversibilities
can be studied. This approach also provides insight
into the performance characteristics of other classes
of real engines where such irreversibilities
may be difficult to model analytically. Thus,
through the lens of thermoelectricity, we can clarify 
the role of certain limiting models which 
ignore specific irreversibilities for the sake
of simplicity. 

Over and above the Constant Properties model, 
we have made two further assumptions: i) a small external irreversibility and ii) equal thermal conductance of heat exchangers at the hot and cold junctions. An advantage of this simplification is
that the power output of a TEG is still a quadratic function
of the electric current---similar to the case of
no external irreversibility. However, the heat fluxes
also depend on the current cubically, which
implies that the model is beyond the linear-irreversible
regime.
We derived a general expression for the efficiency at maximum power
which is determined by the ratio $k$ of internal to external
thermal conductance, apart from Carnot efficiency
and the figure of merit $z$. The SEI model yields the operation
of an engine only in the
regime $k< 1/2$.
Similarly, we have extended the analysis to thermoelectric refrigerators where the efficiency at the maximum cooling power is shown to depend on the Carnot coefficient, the figure of merit $\mathtt{z}$ and the ratio $k$.
The condition $k < 1/2$ is also relevant in the case
of the TER model.
Earlier studies showed that the cooling power is optimizable
only for the exoreversible model. Our analysis extends it into
the regime of small external irreversibility.
In the present model, the limit of a vanishing
external irreversibility can be looked upon as
the smallness of $K_0$ in comparison with $K$ ($K_0 \ll K$).

There are models in literature which report EMP in
the non-linear response regime
\cite{Birjukov2008, AJM2008, SchmiedlSeifert2008, Tu2008,
Esposito2009, esposito2009universality, izumida2008epl,
esposito2010efficiency} beyond the CA value.
The minimally
nonlinear irreversible model \cite{Izumida2012, Izumida2015}
introduces an asymmetric dissipative term in the thermal fluxes.
This term can aid to increase the EMP beyond the
linear response bound.
Such efficiencies are also obtained in the thermoelectric models
based on functionally graded TEMs where an asymmetry
in the Joule heating term can be engineered in the
thermal fluxes \cite{Jasleen2019}. Similarly, an
upper bound of $\eta_{\rm C}/(2-\eta_{\rm C})$ can be derived
for EMP in the tight-coupling approximation.
In this paper, we observe that Joule dissipation also becomes
asymmetric in the presence of external dissipation,
which leads to more heat being dissipated into the hot reservoir
than the cold reservoir. Further, we have considered heat leaks as well as
 non-ideal, though symmetric, thermal contacts.
 The power expression is a quadratic function
 of the current (`useful' flux)
 as in the minimally nonlinear model, but
 the presence of cubic terms in the thermal fluxes
 makes our model go beyond the minimal model.
We have also mapped the TEG in the endoreversible
approximation to the Feynman's ratchet in the
nonlinear, high-temperatures regime and explained the
occurence of the common expression for EMP in these two
models. This provides an alternative perspective to the recent proposal
of Ref. \cite{Apertet_FR2014} where the mapping was
made to the exoreversible limit.

It is hoped that the SEI model may provide  practical
indicators to benchmark the performance of
thermoelectric devices where the heat exchangers'
thermal conductance is much larger than that of the
thermoelectric material.
The present study also builds a  bridge between simpler,
finite-time thermodynamic models and realistic
irreversible machines, offering an insight into the interplay between
external and internal irreversibilities which may guide their practical design.

%  Also, the relevance of minimal model in the context
%  of thermoelectric devices has been questioned in
%  \cite{Apertet2013}. As shown in Ref. \cite{Jasleen2019},
%  the asymmetric dissipation can be treated in thermoelectric
%  devices, say
%  by considering spatially dependent heat conductivity
%  for a TEM.

\section*{Acknowledgments}
RC acknowledges the grant of research fellowship
from Indian Institute of Science Education and Research
Mohali.


%\bibliography{mybib}
\end{document}